\documentclass[sigconf]{acmart}

\AtBeginDocument{
  }

\renewcommand\footnotetextcopyrightpermission[1]{}
\begin{document} 

\title{MoodSmith: Enabling Mood-Consistent Multimedia for AI-Generated Advocacy Campaigns}

\author{Samia Menon}
\affiliation{
  \institution{Columbia University}
  \city{\normalsize New York}
  \state{\normalsize NY}
  \country{\normalsize USA}
}
\email{sm4788@columbia.edu}

\author{Sitong Wang}
\affiliation{
  \institution{Columbia University}
  \city{\normalsize New York}
  \state{\normalsize NY}
  \country{\normalsize USA}
}
\email{sw3504@columbia.edu}

\author{Lydia B. Chilton}
\affiliation{
  \institution{Columbia University}
  \city{\normalsize New York}
  \state{\normalsize NY}
  \country{\normalsize USA}
}
\email{chilton@cs.columbia.edu}

\renewcommand{\shortauthors}{Menon, et al.}

\begin{abstract}
Emotion is vital to information and message processing, playing a key role in attitude formation. 
Consequently, creating a mood that evokes an emotional response is essential to any compelling piece of outreach communication. 
Many nonprofits and charities, despite having established messages, face challenges in creating advocacy campaign videos for social media. 
It requires significant creative and cognitive efforts to ensure that videos achieve the desired mood across multiple dimensions: script, visuals, and audio.
We introduce MoodSmith, an AI-powered system that helps users explore mood possibilities for their message and create advocacy campaigns that are mood-consistent across dimensions. 
To achieve this, MoodSmith uses emotive language and plotlines for scripts, artistic style and color palette for visuals, and positivity and energy for audio. 
Our studies show that MoodSmith can effectively achieve a variety of moods, and the produced videos are consistent across media dimensions.

\end{abstract}



\begin{teaserfigure}
\centering
\includegraphics[width=1\textwidth]{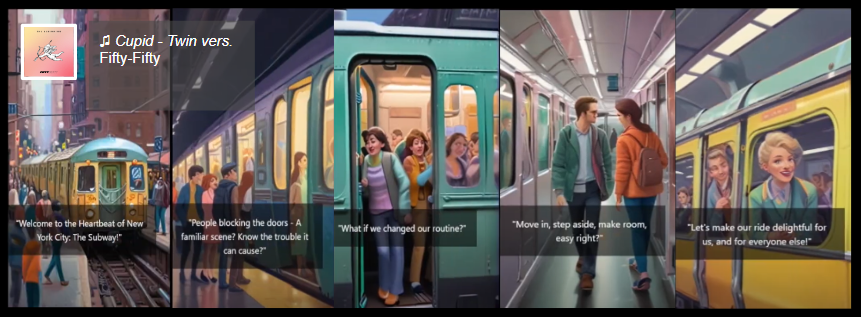}
\caption{A short video advocacy campaign created with MoodSmith encourages New York Subway riders to move all the way in and away from the doors to prevent hazards and delays when boarding and exiting trains.}
\label{fig:teaser}
\end{teaserfigure}

\maketitle
\pagestyle{plain} 

\section{Introduction}
\begin{figure*}
    \centering
    \includegraphics[width=1\linewidth]{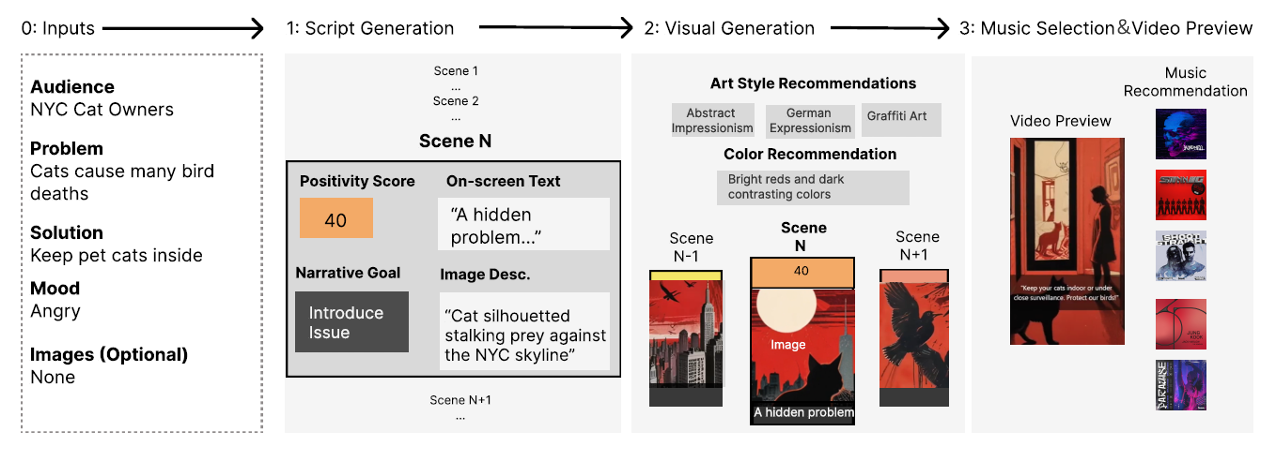}
    \caption{MoodSmith supports users in creating short advocacy campaign videos for a given message and target mood. It consists of three stages: Script Generation, Visual Generation, Music Selection and Video Preview.}
    \label{fig:system_diagram}
\end{figure*}

Emotion is vital to information and message processing, playing a key role in attitude formation~\cite{barrett2016works}. 
Consequently, creating a mood that evokes an emotional response is essential to any compelling piece of outreach communication. 
Public Service Announcements (PSAs), aiming to impact public behavior and awareness, often utilize mood to create persuasive and impactful communications~\cite{dillard2000affect}. 
Different emotional appeals can lead to varied reactions from the audience, with moods like fear, sadness, disgust, and hope each playing a unique role in motivating behavioral change.
For example, while fear can prompt reconsideration of risky behaviors, hope may foster a commitment to behavior change~\cite{dillard2006persuasive}.

Nonprofits and charities often have messages they want to share and would like to use short social media videos for outreach.
However, they face challenges in creating advocacy campaign videos that achieve a target mood. 
Ensuring consistency across various dimensions---such as the script, visuals, and audio---is a challenging task that carries a significant creative and cognitive load.
Achieving a consistent mood demands not just initial concept alignment but also careful attention to detail throughout the process. 
Every component, from the script’s language choice and the visual color scheme to the background music's tone, must work together to effectively evoke the desired emotional response.

Generative AI has the potential to help explore multiple moods and achieve the target mood for a given message.
Incorporating mood-relevant stylistic parameters during AI creation helps.
Specifically, we can utilize emotive language and plotlines for the script, artistic style and color palette for visuals, and positivity and energy for audio. 
Generative AI can provide suggestions, but it faces challenges in maintaining coherence and accuracy.
Humans must be involved in the process to guide and edit the generations.

We present MoodSmith, an AI-powered system that helps users explore mood possibilities for a given message and create advocacy campaigns that are mood-consistent across each dimension. 
By leveraging key video storytelling tools, such as narrative structure, evocative imagery, and music, across three stages: Script Generation, Visual Generation, Music Selection and Video Preview, users can explore and achieve a variety of moods to deliver their message.

This work provides the following contributions:
\begin{itemize}\itemsep0em
    \item A structured prompting technique to produce text, images, and music to achieve a consistent mood for short videos. 
    \item MoodSmith, an interactive human-AI workflow to help non-profits and charities create advocacy campaigns.
    \item Studies showing that MoodSmith can effectively achieve a diversity of moods, and the produced videos are consistent across media dimensions. 
\end{itemize}

\section{Related Work}

Emotion plays a vital role in human information processing and significantly influences the actions humans undertake~\cite{barrett2016works}. 
The circumplex model of affect~\cite{posner2005circumplex} is a fundamental framework within psychology that elucidates critical dimensions of emotion---valence and arousal. 
While valence captures how positive or negative the emotion is, arousal measures the degree of activation that the emotion entails. 
In this model, the two dimensions help define a plane that expresses nuances in how emotions differ. This helps explain how different emotions impact human behavior. 
For example, fear is characterized by negative valence and high arousal, compelling people to take action, such as ``fight or flight''. 
In contrast, the experience of depression, though also marked by negative valence, is associated with low arousal, resulting in a lack of motivation to engage in escape behavior. 
Positive emotions can also have different arousals.
Calm or contentment is positive but has low arousal.
Excitement is positive but has high arousal---leading to actions like starting a company or donating money.

In the field of literature and cinematic arts, creators often evoke feelings or emotions in the audience by carefully creating the mood of their work~\cite{hogan2011affective,phillips2009film}. 
Creating a mood is a complex art that requires the careful arrangement of various elements to elicit the desired emotional response. 
For example, suspense in a drama can be achieved through dim lighting, tense music, tight frames in a movie, or ominous language and unresolved conflict in literature. 
Creating a particular mood is a creative challenge that requires a nuanced and comprehensive approach.

The strategic use of moods in PSAs has been explored in the field of communication. 
Research has delineated how distinct emotions, when evoked through PSAs, can facilitate different types of reactions and actions among viewers~\cite{dillard2000affect}. 
For example, in cancer prevention PSAs, fear highlights risks, sadness encourages reflection on the consequences of inaction, disgust creates negative associations with risky behaviors, and hope inspires action seen as unattainable~\cite{dillard2006persuasive}. 
The valence of moods influences the effectiveness of PSAs. 
For example, a study on road safety PSAs highlights the effectiveness of positive emotional appeals in promoting health behaviors ~\cite{lewis2007promoting}. 
Levels of arousal also impact the effectiveness. 
For example, experiments have found that smokers tend to dismiss anti-smoking PSAs that are overly fear-inducing ~\cite{keller1999converting}.
It is, therefore, important for PSA creators to explore a variety of moods and craft content tailored to elicit the intended mood.

AI is a powerful creative tool that can manipulate narrative elements to impact emotion and mood.
Many HCI works have explored such AI applications. 
For example, AngleKindling~\cite{anglekindling} aids in finding news angles from press releases, enabling journalists to transform positive press leads into negative controversies; 
Talebrush~\cite{talebrush} and Dramatron~\cite{dramatron} construct a story through a narrative arc, detailing the main character’s ups and downs; 
ReelFramer~\cite{reelframer} assists in reframing a news article into a positive and upbeat social media reel. 
In terms of visuals, MayAI~\cite{mayai} helps users create a mood board by suggesting images based on semantics, color, saturation, and brightness; 
EmoG~\cite{emog} aids in generating character designs with varying emotional expressions and intensities. 
For audio, Cococo~\cite{cococo} supports music generation according to emotional quality, such as minor vs major tones; 
An automated generation system~\cite{rubin2014generating} creates emotionally relevant music scores for the speech of audio stories, spanning four basic emotions: happy, sad, nervous, and calm. 
MoodSmith aims to establish a human-AI workflow to create an advocacy campaign that amplifies a message in three formats: script, visuals, and audio. 
It helps users explore a diverse set of moods and ensure mood consistency within and across the modals.

\section{MoodSmith System}

A popular type of social media campaign is the slideshow-like animated reels~\cite{Cavender2023}. 
These reels are heavily driven by images, with scripted texts on screen and accompanied by trending music. 
We explored using generative AI to craft texts, images, and audio that work together to create such reels to communicate the public service message.

We introduce MoodSmith (see Figure \ref{fig:system_diagram}), a three-stage AI-powered system that helps users create short-form advocacy campaign videos for an input message.  
The stages include: 
1) \textit{Script Generation}, which primarily focuses on refining the mood of the video’s narratives,
2) \textit{Visual Generation},  which allows the user to generate images to achieve a particular mood, leveraging art style and color suggesions, and
3) \textit{Music Selection and Video Preview}, which provides music options to align with the mood of the rest of the media.

The system was implemented in React, and it is powered by GPT-4 (text2text, img2text)~\cite{gpt4}, Stable Diffusion (text2img, img2img)~\cite{stable_diffusion}, Sentiment (AFINN-based sentiment analysis)\footnote{\url{https://github.com/thisandagain/sentiment}}, string-similarity-js (Sørensen–Dice coefficient-based string comparison)\footnote{\url{https://www.npmjs.com/package/string-similarity-js}}, and the Spotify Web API\footnote{\url{https://developer.spotify.com/documentation/web-api}}. 
The prompts used can be found in the Appendix.   
To demonstrate the workflow, we describe an interaction a user may have with the system.

\subsection{Initial Input}
To begin creating an advocacy campaign, the user inputs 1) audience, 2) problem, 3) actionable solution, 4) desired mood.
These inputs were chosen to align with the goals of many advocacy campaigns---to motivate the audience to action. 
For example, if a conservation organization in NYC wants to inform people to keep their cats at home to preserve birds, their inputs can be as follows: 1) audience: cat owners in NYC; 2) problem: pet cats cause the loss of 2.4 billion birds each year in the US; 3) actionable solution: keep pet cats indoors; 4) mood: calm.

Note that in this step, there is also an optional choice to upload images to incorporate. 
If the user wishes to integrate existing media or is particular about image accuracy, such as a conservation organization that wants to ensure an accurate representation of a specific species, they can choose to upload their own images. 
For each user-uploaded image, GPT-Vision provides a brief description of the image content that the user can adjust for accuracy. 
This description is used to integrate the uploaded images in an appropriate context later in the video-making process.

\subsection {Stage 1: Script Generation} 
\begin{figure}
    \centering
    \includegraphics[width=0.9\linewidth]{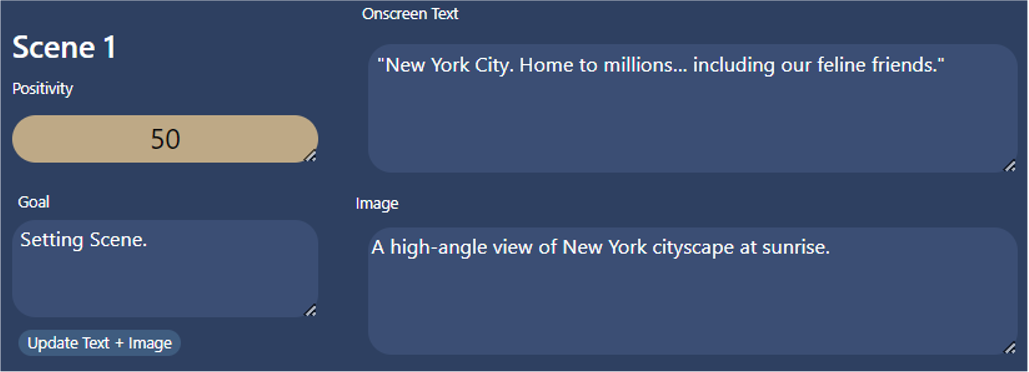}
    \caption{An example scene in Stage 1: Script Generation.}
    \label{fig:example-scene}
\end{figure}

From the inputted information, the system generates a script using GPT-4. 
The script is composed of a series of scenes.
Every scene includes text content, an image description, and a suggested narrative goal to achieve (see Figure \ref{fig:example-scene}).
Additionally, each scene is assigned an initial positivity score ranging from 0 (most negative) to 100 (most positive). 
Here, positivity refers to the ``valence'' value of each individual scene.
The positivity score is calculated through sentiment analysis on both the text and the image description. 

The Script Generation stage is used as an establishing step: the user ensures that the foundational structure of their video---the narrative established by the on-screen text and image descriptions---aligns with their desired mood, before generating visual and aural components. 
The user can evaluate the positivity level of the script using the color-coded positivity scores. 
The narrative goals summarize the objective of each scene, allowing the user to quickly determine whether or not each beat of the narrative arc, or ``plot'', of the video aligns with their desired mood. 
For each scene, the user can edit the positivity score and narrative goal before regenerating the text content and image description. 
For example, if the user attempts to achieve a hopeful mood, and the final scene generated has a positivity score of 70 and a narrative goal of ``Describing next steps'', the user may adjust the positivity score to 90 and narrative goal to ``Imagining a brighter tomorrow'', to further underline the intended mood. 
These two levels of customization influence 1) the positivity of the word choice and 2) the story structure, two key stylistic elements that influence mood of text. 

User's uploaded images (if any) are incorporated into the script: the description of each uploaded image is compared to each scene's generated image description using string-similarity-js. 
In the scene with the closest match, the scene image description is replaced with the uploaded image description. 
The replacement is done so the user is aware where their uploaded images will appear in the next stage. 

\subsection{Stage 2: Visual Generation}
\begin{figure}
    \includegraphics[width=1\linewidth]{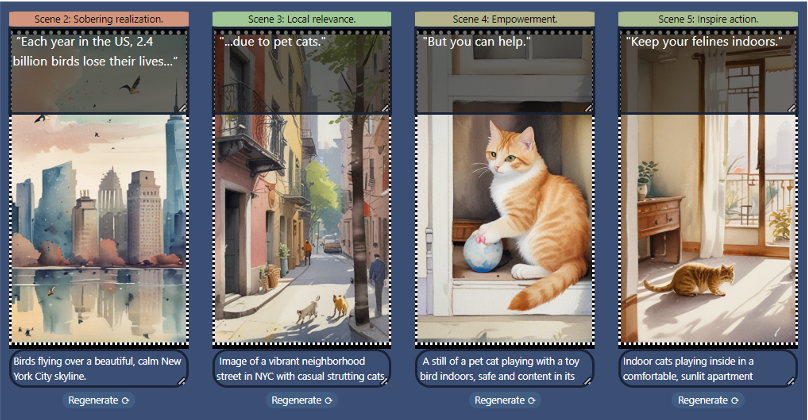}
    \caption{An example sequence of images in Stage 2: Visual Generation. The mood is ``calm''. The art style is ``Chinese Watercolor Painting'' with ``soft, soothing colors''. }
    \label{fig:example-images}
\end{figure}

\begin{figure}
    \centering
    \includegraphics[width=1\linewidth]{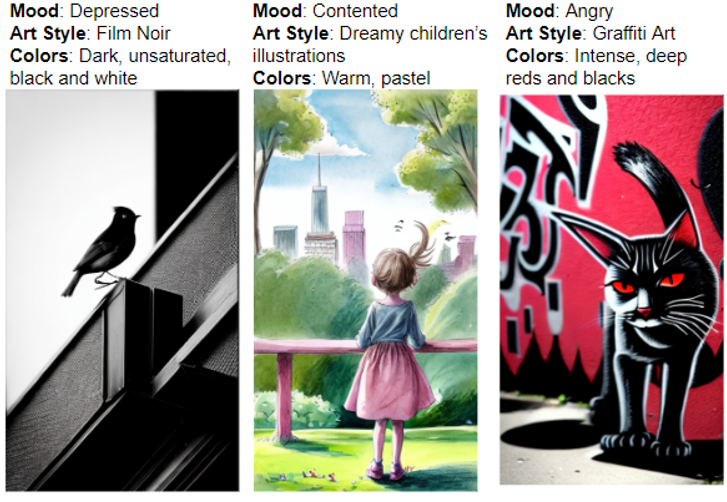}
    \caption{Examples of art style and color recommendations for three moods: depressed, contented, and angry.}
    \label{fig:artstyle}
\end{figure}

Once the user is satisfied with the script, the system helps them generate a sequence of images to convey the scenes and the mood visually (see Figure \ref{fig:example-images}).

The mood of the images is controlled through the choice of art style and color. 
Although subjective, many art styles contain certain traits that tend to evoke specific emotions. 
For example, the stark, abstract shapes of German Expressionism are rife with tension and discomfort. 
Similarly, colors is vital in emotional perception---while bright, saturated colors carry high levels of energy, pastels bring a sense of calm. 
These two dimensions can transform the mood of images and are important tools for the user.
The system suggests three art styles and a color palette to use for the images (see examples in Figure \ref{fig:artstyle}).
It assesses the script's sentiment---from strongly negative to strongly positive---based on the average positivity score of the scenes.
Based on this score and the inputted mood, the system then asks suggestions from GPT-4.
This is to ensure the system accounts for the user's original vision and any adjustments they made in the previous Stage Generation stage.

The user can select or regenerate art style and color suggestions. 
After making a selection, the system generates images for each scene. 
These images are created by inputting the chosen art style and color, along with corresponding image descriptions (determined during the previous Script Generation stage), into Stable Diffusion’s text-to-image endpoint. 
To ensure that the editing process does not ``erase'' their inclusion, the uploaded images (if any) are reintegrated into the generation.
This is again achieved by using string similarity on the descriptions to determine their placement.
The user-uploaded images are then restyled using the image-to-image endpoint. 
Each image is then rendered in a 19.5:9 aspect ratio to match a mobile video format, with each scene’s on-screen text overlaid on it. 
The user can click on each image to see three other generations to choose from to explore other options. 

Each scene's visuals are topped with a color-coded label to indicate the positivity score and narrative goal that was established in the previous Script Generation stage. 
This allows users to quickly compare the visuals to the established positivity and narrative goal of the scene to evaluate the mood consistency between dimensions. 
At any point, the user can generate and assess new art style and color recommendations. 
The user also has the option to edit individual image descriptions and regenerate or restyle the images (if they want to preserve the core content of the images).

\subsection{Stage 3: Music Selection and Video Preview}
\begin{figure}
    \centering
    \includegraphics[width=0.99\linewidth]{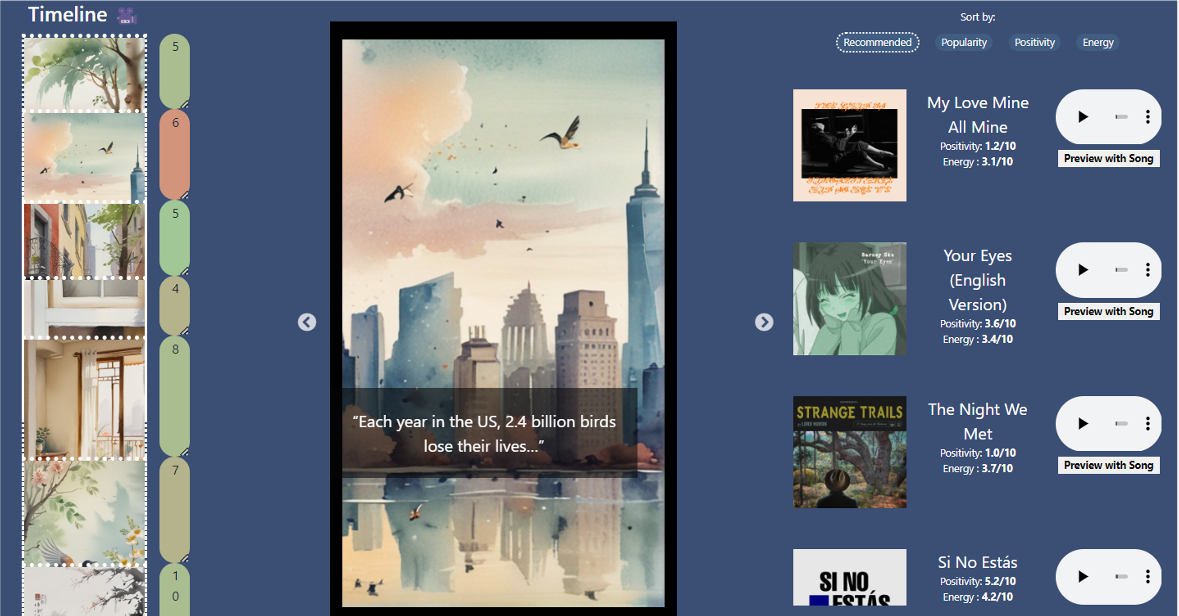}
    \caption{An example preview in Stage 3: Music Selection and Video Preview.}
    \label{fig:video-preview}
\end{figure}

In the final stage, the user can choose music that is consistent with the mood and popular on social media. They can also preview the final video alongside the selected music (see Figure \ref{fig:video-preview}). 

The system recommends songs that are most consistent with the mood of the rest of media from a popular song collection. 
The song collection is retrieved from a TikTok Top Hits playlist\footnote{\url{https://open.spotify.com/playlist/65LdqYCLcsV0lJoxpeQ6fW}} via the Spotify API.
The API provides a ``valence'' value and an ``energy'' value for each song. 
The system recommends songs that have the smallest distance with the target score of ``valence'' and ``energy''. 
The target ``valence'' value is the average positivity score of all the scenes (normalized to a value between 0-1).
The target ``energy'' value is the original inputted mood's energy value evaluated by GPT-4, between 0 (extremely calm) and 1 (extremely excited).
By using the positivity score established in the previous stages, and the mood used throughout, the system prioritizes the consistency of the suggested song with other video elements.
The user also has the option to rank the songs based on popularity, where the system organizes the songs from the highest Spotify popularity score to the lowest. 

After selecting the music, the user can click the ``Preview with Audio'' button to view their video. 
Each scene's visuals will display for a period of time recommended by GPT-4.
As always, these values are editable, and can be accessed in the Timeline display. 
The Timeline provides a quick overview of the video, displaying each scene’s image, duration, and positivity score, providing the user a basis for evaluation. 
It is placed to the left of the Video Preview for easy reference. 

\section{Technical Evaluation}

To measure the effectiveness of MoodSmith's workflow, we conducted a technical evaluation that focuses on three dimensions: 
1) \textit{accuracy}: the ability to achieve the mood specified in the input; 
2) \textit{consistency}: the ability to create videos that are consistent in mood across dimensions (text, imagery, and audio); 
3) \textit{clarity}: the ability to generate videos that achieve moods clear to audience. 
We compared the latter two dimensions with a baseline workflow that did not include the key mood adjustment features of MoodSmith.

\subsection{Data Collection}

\begin{figure}
    \centering
    \includegraphics[width=0.75\linewidth]{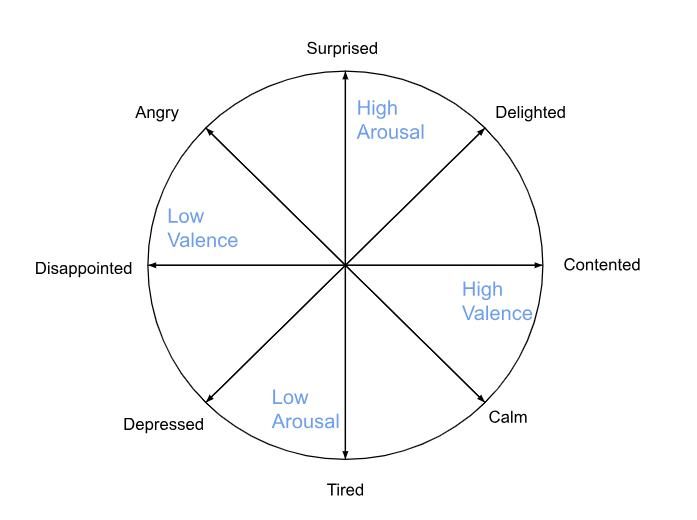}
    \caption{Moods used in our technical evaluation.}
    \label{fig:emotion-chart}
\end{figure}

For our technical evaluation, we selected 8 moods, ensuring that each is distinct and spanned evenly across the valence-arousal spectrum, as illustrated in Figure \ref{fig:emotion-chart}.

We generated 16 videos with the key mood adjustment features of MoodSmith.
We call these videos ``with-mood'' generations.
8 videos used one message (M1) as the input, and the other 8 used another (M2).  
(See content of M1 and M2 in Appendix.)
Within each message set, 8 moods were represented (one for each video).  
During the generation, choices of art style and color were taken from the first set of suggestions by the system, and the choice of music was taken from the top 8 songs recommended by the system. 

As the baseline, we generated 16 videos (8 M1, 8 M2, one mood for each) without the key mood adjustment features of the system. 
We call these videos ``without-mood'' generations.
The baseline workflow is similar to that of MoodSmith in all possible ways, except in baseline, mood was not incorporated in any of the prompts inputted to the LLM (see Appendix). 
During the generation, art style and color recommendations were removed, and the final song was chosen at random.  
For all videos, on-screen text and image descriptions were minimally edited for accuracy and coherence. 

\subsection{Participants and Procedure}
To evaluate the generated videos, we recruited four annotators (average age=22.3; 4 male) through word-of-mouth.
The annotators were active on social media and consumed video content daily. 
Each was asked to complete a survey that covers half of the generated videos. 
Each video is annotated by two different annotators.
The order of the videos in the survey were randomized in terms of the with- and without-mood conditions.
In the survey, annotators were asked to watch the videos and identify the mood of the \textit{text} of the video, the \textit{imagery} of the video, the \textit{music} of the video, and the \textit{overall} video. 
To identify the mood, annotators were given multiple-choice options of the 8 moods and ``unclear''. 
Annotators were compensated \$25.
Their answers were used to inform the findings below.

\subsection{Findings}
\subsubsection{MoodSmith produced videos that were reasonably \textit{accurate} to the intended mood.}
For each video, two annotators provided the \textit{overall} mood identifications. 
From the annotations, we calculate the exact match accuracy, valance match accuracy and arousal match accuracy.
They are determined by whether at least one annotator correctly identifies the video's mood as the target mood, a mood with the same valence, or a mood with the same arousal level.
For example, if the target mood was ``calm'', but an annotator selected ``tired'' or ``depressed'', it would qualify as an accurate response under an arousal match because of the shared ``low arousal'' between the moods; however, these would not qualify under exact match.  If the annotator selected ``contented'' or ``delighted'', it would qualify under a valence match.

Overall, 37.5\% of the videos were accurately categorized under an exact match.
75.0\% and 56.2\% of the videos were accurate under a valence match and an arousal match, respectively.
Choosing one mood from eight possible options (plus ``unclear'') is a challenging task for annotators who are not familiar with the valence-arousal theory. 
Understandings and associations with specific moods are highly personal and subjective. 
For instance, some annotators associated ``tired'' (defined to have neutral valence) with a significantly more negative valence, or ``content'' (defined to have neutral arousal) with lower arousal. 
This highlights the subjective nature of mood perception and the complexity of accurately categorizing mood-based content. 
Considering these challenges, the match accuracy results demonstrate that MoodSmith can reasonably capture the target mood.

\subsubsection{MoodSmith produced videos that were more \textit{mood-consistent} across channels than the baseline.}
For each video, we calculated a consistency score as follows: for each annotator and for each channel's mood assignment (text, imagery, audio), one point was given for every exact match with the overall video mood assignment; 
a half point was given for a mood assignment that shared the same valence or arousal as the overall video mood assignment. The highest possible consistency score is 3, representing an exact match for all three dimensions.

The average consistency score for a without-mood video was 1.02 (SD=1.05), indicating that the annotator's final choice in determining the video mood was largely informed by the mood of only one channel.  
The average consistency score for a with-mood video was 1.61 (SD=1.02), showing that the with-mood videos often have at least two mood-consistent channels. Under a t-test, this result is shown to be highly significant, with a p value of 0.03 (<0.05). 

\subsubsection{MoodSmith produced videos that achieved greater \textit{clarity} of mood than the baseline.}
Overall, the without-mood videos had 14 ``unclear'' identifications, which account for 43.8\% of the total. 
Meanwhile, the with-mood videos recorded 7 ``unclear'' identifications, representing 21.9\% of their total. 
This indicates that videos produced with key mood adjustment features of MoodSmith were more likely to convey a clear mood than the baseline.
The 50\% reduction in ``unclear'' identifications illustrates the effectiveness of MoodSmith in enhancing the emotional conveyance of videos.
Through its key mood adjustment features, MoodSmith has demonstrated its capability to refine and intensify mood portrayal, making videos more clear and understandable from an emotional perspective.

\section{User Study}
Our technical evaluation has demonstrated MoodSmith's capability when automated. 
We believe that with human creators involved in the process, the results will be even better.
To determine MoodSmith’s ability to support users in creating advocacy campaigns, we conducted a user study involving five participants with experience working in nonprofit organizations.
Each was asked to create two videos regarding an issue they felt passionate about with two distinct moods.

\subsection{Participants and Procedure}
We recruited five participants (average age=23.0; three female, two male) with a background working in non-profit organizations in either a professional or volunteer capacity. 
Two participants held full-time positions at a non-profit organization at the time of the study, and four participants reported having over three years of highly involved experience at a non-profit organization. 
In terms of their experience of using generative AI, one participant reported working with generative AI tools daily, two working with them a few times a month, one working with them a few times a year, and one had never used them before.

Each participant had the goal of creating two videos with two distinct moods on a topic of their choice.  
For the first video, the participant received instructions from the facilitator regarding how to use the system. 
Along with inputting their own desired message, they were shown a few examples of moods (happy, sad, calm, disappointed, angry) and then chose a mood of their own. 
Once they were satisfied with their video generation, we conducted a semi-structured interview regarding their experience. 
For the second video, the participant was instructed to use the exactly same message as the first video, but to choose a new mood they believed was substantially different from their first choice. 
Once they were satisfied with their second video generation, another semi-structured interview was conducted, and the user filled out a survey to reflect on their outputs and experience with the system.  
The survey contains questions of helpfulness scores of specific features, NASA TLX~\cite{nasa-tlx} and user's scores of their final outputs.
The study was around 45 minutes.
Participants were compensated \$25.

\subsection{Findings}
\subsubsection{MoodSmith allowed participants to easily explore a diverse range of moods and create videos to achieve the moods.}

During the study, participants explored a variety of moods, including desperate, excited, content, depressed, energetic, hopeful, happy, and tranquil. 
When asked how much they agreed with the statement ``It was easy for me to explore many different moods for my video without tedious, repetitive interaction.'', participants provided an average score of 6.0 out of 7 (SD=1.10).  
P4, who gave a score of 4, agreed that it was \textit{``relatively easy to change the mood''} by altering the initial input; however, they had to \textit{``continue making adjustments to match the intended mood.''} 
Participants found the system easy to use---as P5 commented \textit{``The linearity of the process made it very intuitive. It’s one of the most easy-to-use AI systems I’ve worked with.''} 

Specifically, participants found the positivity score and narrative goal features very helpful in supporting them to achieve their creative goal. 
The positivity score feature achieved a score of 5.8 out of 7 (SD=0.75) in terms of helpfulness.
All users referenced positivity scores as they evaluated the quality of the script, visuals and music---P1 commented \textit{``I find them [positivity scores] very helpful as a reference point.''}, and noted that they were helpful in evaluating whether or not the overall video was achieving the desired mood. 
The narrative goal feature received an average helpfulness score of 6.6 out of 7 (SD=0.49).  
Four out of five participants mentioned they used the narrative goal as a tool to clarify the motivation of specific scenes in the script. 
They were impressed with the accuracy of the generated results and how it integrated into the script. 
The specific helpfulness of other features related to the image and audio dimensions will be discussed in the next section. 

With the system, users were able to explore two moods with relatively little effort (3.0/7, SD=0.98) and mental demand (2.2/7, SD=0.75) while completing their task of creating two advocacy campaign videos with an average self-assessed success of 4.8/7 (SD=0.98).
Four out of five participants assessed their success as 5 or above out of 7. 
Next, we delve deeper into users' perceived success in achieving their intended mood across different dimensions of the video.

\begin{figure}
    \centering
    \includegraphics[width=1\linewidth]{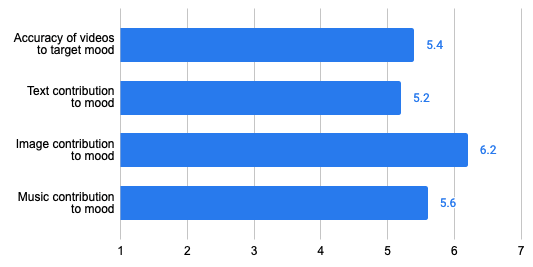}
    \caption{Average scores on mood accuracy and contribution of each dimension to mood from the user study survey. }
    \label{fig:user_study_results}
\end{figure}

\subsubsection{MoodSmith supports users in creating videos that accurately achieved their intended moods and were consistent across text, imagery, and audio.}
When asked how accurately the final videos achieved their desired mood, four participants gave a score of 5 and above. 
The average score, 5.4 (SD=0.80), out of 7, indicates that the final generations were consistent with the users’ intended moods. 
P2, who provided a score of 4/7 on this question, noted that some moods could be achieved more successfully than others, explaining that \textit{``The first mood (contented) was accurately reflected and provoked peaceful feelings, while the second time (desperate) did not convey desperation, but [the second one] did provoke strong feelings in a way the first one did not.''} 
All five participants mentioned that the final videos produced by MoodSmith would be helpful in the advocacy campaign crafting process. 
P5 mentioned \textit{``It would be strong at creating drafts on the fly that accurately capture the mood and narrative of an advocacy campaign, while roughly approximating the images.''} 
Participants agree that human edits are still needed to add the personal touch to the campaign.

When asked how significantly each dimension (text, imagery, and audio) contributed to the mood, all were shown to be positive contributors. 
Text received an average score of 5.2 (SD=1.33), imagery a score of 6.2 (SD=0.75), and music a score of 5.6 (SD=1.85), all out of 7 (see Figure \ref{fig:user_study_results}).
From these scores, we can see that each dimension was consistent with the user’s intended mood.

Text provided the necessary structure for the other dimensions’ expression. 
P2 believed that the text contributed to the mood, scoring it a 5/7, but it was \textit{``outshined by what the art's color and style contributed''}. 
P5 agreed with this sentiment, also providing a score of 5/7, explaining that \textit{``text helped with understanding the narrative of the images''}, but \textit{``I don't think the text itself conveys as much to the viewer as opposed to the images and music.''}

Images were the most impactful dimension in influencing mood. 
Participants noted that the art styles were appropriate 
and were key to delivering the video's narrative.
For example, P2  pointed out that \textit{``the color palettes and art styles helped contextualize the captions.''}
P5 also commented \textit{``Imagery really helped to solidify the problem and solution we are looking for in the campaign.''} 
However, existing context and associations with art styles occasionally worked in opposition to the user's messages. 
For example, P4 mentioned the style they chose  (Art Deco) \textit{``ultimately didn't work for the video because it is associated with a certain time period, while the issue the video was about was not relevant to that time period.''} 
P3 also expressed interest in images that could  \textit{``capture a more `serious' atmosphere while still achieving the intended mood''}, suggesting \textit{``there should be a more realistic option available.''} 

Music was an influential but polarizing dimension.
Songs provided in the system were collected from popular TikTok audios, a genre that may not equally represent the spectrum of mood.  
P2, who scored a 4/7 on accuracy to mood and a 2/7 on music’s specific contribution to mood, noted: \textit{``I would have preferred to have music options with no sound, or songs that were not super popular because it distracts from the message [of the video], especially if I had already heard the song before.''} 
Participants who provided a high score for the music’s contribution echoed this sentiment: \textit{``The music selection was very accurate for energetic, and decent for tranquil...maybe there are fewer tranquil TikTok famous songs?''} (P1).  
Due to the above reasons, when asked if the recommended music was accurate, the results were split with an average of 4.2 (SD=0.98). 
Still, most users considered music one of the most influential factors in mood with four out of five participants scoring it a 6 or above. 
During the study, many noted that the addition of music feels like it can completely change the mood of the video. 
As P5 pointed out, \textit{``When the music works for the video...it contributes significantly to the mood.''}

\subsubsection{Suggestions on future improvement}
During the interview, participants provided suggestions for future improvement of MoodSmith.
First, as P4 suggested, further investigation is needed to adapt the art style recommendations to address contradictions based on time periods, such as those related to technology or ethnic representation. 
Additionally, P1, P2 and P5 mentioned that a broader range of music styles would help achieve a wider variety of moods and provide the correct ambiance for more professional settings. 
P3 and P5 have also suggested exploring mechanisms to allow for \textit{shifting} moods within a single video, which presents a significant opportunity for future development.
Implementing this could involve developing dynamic transition tools that blend visuals and audio, enabling creators to evolve the overall mood naturally as the narrative progresses.

\section{Discussion}
MoodSmith helps users create short video advocacy campaigns that are consistent in mood across various media dimensions. 
Our studies have shown that MoodSmith enables participants to explore a variety of moods easily and create videos that effectively convey these moods. 
Emotional narratives are powerful in outreach communication. 
However, non-profit organizations and charities often need to rely on their social media teams (some may not even have such support) to create these compelling outreach artifacts. 
With the advent of generative AI, it is possible to democratize such creation among people who did not have a design background. 
Generative AI  serves as an effective prototyping tool---it can generate numerous ideas and drafts in a very short time.
This opens up new avenues for storytelling and campaign messaging, particularly for smaller organizations with limited resources. 
By leveraging generative AI, they can craft compelling campaigns that resonate with their audience, ultimately enhancing their impact and outreach.
However, AI responses are not always accurate or useful, the challenge lies in how to build workflows around generative AI that people find useful and powerful.

The exploration of mood in the development of creative artifacts represents a promising area within the HCI field. 
As technologies continue to evolve, there is an increasing opportunity to better understand and predict the emotional impact of various forms of media. 
This understanding can then be applied to enhance the creative process, not only in short video production but across a wide array of content creation, such as digital storytelling and interactive design. 
Future research may focus on the development of tools that empower users to integrate complex emotional narratives into their work. 
These tools could incorporate adaptive interfaces that respond to the creator's emotional intent, or even biofeedback mechanisms to tailor content in real-time based on the audience's emotional responses. 
The integration of mood exploration into the content creation tools holds the potential to transform the way stories are told and experiences are crafted, fostering a richer and more empathetic connection between creators and their audiences.

\section{Conclusion}
This paper presents MoodSmith, an AI-powered interactive system that enables users to create short video advocacy campaigns that are mood-consistent in text, visuals and audio.
The system leverages key video storytelling tools,
such as emotive language and plotlines for text, artistic style and color palette for visuals, and positivity and energy for audio. 
Our studies show that with MoodSmith, users can explore and achieve a variety of moods to effectively deliver their message.
We discuss future work on mood exploration in the development of creative artifacts.

\bibliographystyle{ACM-Reference-Format}
\bibliography{sample-base}


\begin{thebibliography}{20}


\ifx \showCODEN    \undefined \def \showCODEN     #1{\unskip}     \fi
\ifx \showDOI      \undefined \def \showDOI       #1{#1}\fi
\ifx \showISBNx    \undefined \def \showISBNx     #1{\unskip}     \fi
\ifx \showISBNxiii \undefined \def \showISBNxiii  #1{\unskip}     \fi
\ifx \showISSN     \undefined \def \showISSN      #1{\unskip}     \fi
\ifx \showLCCN     \undefined \def \showLCCN      #1{\unskip}     \fi
\ifx \shownote     \undefined \def \shownote      #1{#1}          \fi
\ifx \showarticletitle \undefined \def \showarticletitle #1{#1}   \fi
\ifx \showURL      \undefined \def \showURL       {\relax}        \fi
\providecommand\bibfield[2]{#2}
\providecommand\bibinfo[2]{#2}
\providecommand\natexlab[1]{#1}
\providecommand\showeprint[2][]{arXiv:#2}

\bibitem[Barrett(2016)]%
        {barrett2016works}
\bibfield{author}{\bibinfo{person}{Paul~H Barrett}.} \bibinfo{year}{2016}\natexlab{}.
\newblock \bibinfo{booktitle}{\emph{The Works of Charles Darwin: Vol 23: The Expression of the Emotions in Man and Animals}}.
\newblock \bibinfo{publisher}{Routledge}.
\newblock


\bibitem[Cavender(2023)]%
        {Cavender2023}
\bibfield{author}{\bibinfo{person}{E. Cavender}.} \bibinfo{year}{2023}\natexlab{}.
\newblock \bibinfo{title}{Why is everyone on TikTok obsessed with slideshows?}
\newblock
\newblock
\urldef\tempurl%
\url{https://mashable.com/article/tiktok-photo-mode-day-in-the-life}
\showURL{%
\tempurl}
\newblock
\shownote{Accessed: 2023-04}.


\bibitem[Chung et~al\mbox{.}(2022)]%
        {talebrush}
\bibfield{author}{\bibinfo{person}{John Joon~Young Chung}, \bibinfo{person}{Wooseok Kim}, \bibinfo{person}{Kang~Min Yoo}, \bibinfo{person}{Hwaran Lee}, \bibinfo{person}{Eytan Adar}, {and} \bibinfo{person}{Minsuk Chang}.} \bibinfo{year}{2022}\natexlab{}.
\newblock \showarticletitle{TaleBrush: Sketching stories with generative pretrained language models}. In \bibinfo{booktitle}{\emph{Proceedings of the 2022 CHI Conference on Human Factors in Computing Systems}}. \bibinfo{pages}{1--19}.
\newblock


\bibitem[Dillard and Nabi(2006)]%
        {dillard2006persuasive}
\bibfield{author}{\bibinfo{person}{James~Price Dillard} {and} \bibinfo{person}{Robin~L Nabi}.} \bibinfo{year}{2006}\natexlab{}.
\newblock \showarticletitle{The persuasive influence of emotion in cancer prevention and detection messages}.
\newblock \bibinfo{journal}{\emph{Journal of communication}} \bibinfo{volume}{56}, \bibinfo{number}{suppl\_1} (\bibinfo{year}{2006}), \bibinfo{pages}{S123--S139}.
\newblock


\bibitem[Dillard and Peck(2000)]%
        {dillard2000affect}
\bibfield{author}{\bibinfo{person}{James~Price Dillard} {and} \bibinfo{person}{Eugenia Peck}.} \bibinfo{year}{2000}\natexlab{}.
\newblock \showarticletitle{Affect and persuasion: Emotional responses to public service announcements}.
\newblock \bibinfo{journal}{\emph{Communication research}} \bibinfo{volume}{27}, \bibinfo{number}{4} (\bibinfo{year}{2000}), \bibinfo{pages}{461--495}.
\newblock


\bibitem[Hart and Staveland(1988)]%
        {nasa-tlx}
\bibfield{author}{\bibinfo{person}{Sandra~G Hart} {and} \bibinfo{person}{Lowell~E Staveland}.} \bibinfo{year}{1988}\natexlab{}.
\newblock \showarticletitle{Development of NASA-TLX (Task Load Index): Results of empirical and theoretical research}.
\newblock In \bibinfo{booktitle}{\emph{Advances in psychology}}. Vol.~\bibinfo{volume}{52}. \bibinfo{publisher}{Elsevier}, \bibinfo{pages}{139--183}.
\newblock


\bibitem[Hogan(2011)]%
        {hogan2011affective}
\bibfield{author}{\bibinfo{person}{Patrick~Colm Hogan}.} \bibinfo{year}{2011}\natexlab{}.
\newblock \bibinfo{booktitle}{\emph{Affective narratology: The emotional structure of stories}}.
\newblock \bibinfo{publisher}{U of Nebraska Press}.
\newblock


\bibitem[Keller(1999)]%
        {keller1999converting}
\bibfield{author}{\bibinfo{person}{Punam~Anand Keller}.} \bibinfo{year}{1999}\natexlab{}.
\newblock \showarticletitle{Converting the unconverted: the effect of inclination and opportunity to discount health-related fear appeals.}
\newblock \bibinfo{journal}{\emph{Journal of Applied Psychology}} \bibinfo{volume}{84}, \bibinfo{number}{3} (\bibinfo{year}{1999}), \bibinfo{pages}{403}.
\newblock


\bibitem[Koch et~al\mbox{.}(2019)]%
        {mayai}
\bibfield{author}{\bibinfo{person}{Janin Koch}, \bibinfo{person}{Andr{\'e}s Lucero}, \bibinfo{person}{Lena Hegemann}, {and} \bibinfo{person}{Antti Oulasvirta}.} \bibinfo{year}{2019}\natexlab{}.
\newblock \showarticletitle{May AI? Design ideation with cooperative contextual bandits}. In \bibinfo{booktitle}{\emph{Proceedings of the 2019 CHI Conference on Human Factors in Computing Systems}}. \bibinfo{pages}{1--12}.
\newblock


\bibitem[Lewis et~al\mbox{.}(2007)]%
        {lewis2007promoting}
\bibfield{author}{\bibinfo{person}{Ioni~M Lewis}, \bibinfo{person}{Barry Watson}, \bibinfo{person}{Katherine~M White}, {and} \bibinfo{person}{Richard Tay}.} \bibinfo{year}{2007}\natexlab{}.
\newblock \showarticletitle{Promoting public health messages: Should we move beyond fear-evoking appeals in road safety?}
\newblock \bibinfo{journal}{\emph{Qualitative health research}} \bibinfo{volume}{17}, \bibinfo{number}{1} (\bibinfo{year}{2007}), \bibinfo{pages}{61--74}.
\newblock


\bibitem[Louie et~al\mbox{.}(2020)]%
        {cococo}
\bibfield{author}{\bibinfo{person}{Ryan Louie}, \bibinfo{person}{Andy Coenen}, \bibinfo{person}{Cheng~Zhi Huang}, \bibinfo{person}{Michael Terry}, {and} \bibinfo{person}{Carrie~J Cai}.} \bibinfo{year}{2020}\natexlab{}.
\newblock \showarticletitle{Novice-AI music co-creation via AI-steering tools for deep generative models}. In \bibinfo{booktitle}{\emph{Proceedings of the 2020 CHI conference on human factors in computing systems}}. \bibinfo{pages}{1--13}.
\newblock


\bibitem[Mirowski et~al\mbox{.}(2023)]%
        {dramatron}
\bibfield{author}{\bibinfo{person}{Piotr Mirowski}, \bibinfo{person}{Kory~W Mathewson}, \bibinfo{person}{Jaylen Pittman}, {and} \bibinfo{person}{Richard Evans}.} \bibinfo{year}{2023}\natexlab{}.
\newblock \showarticletitle{Co-writing screenplays and theatre scripts with language models: Evaluation by industry professionals}. In \bibinfo{booktitle}{\emph{Proceedings of the 2023 CHI Conference on Human Factors in Computing Systems}}. \bibinfo{pages}{1--34}.
\newblock


\bibitem[OpenAI(2023)]%
        {gpt4}
\bibfield{author}{\bibinfo{person}{OpenAI}.} \bibinfo{year}{2023}\natexlab{}.
\newblock \bibinfo{title}{GPT-4 Technical Report}.
\newblock
\newblock
\showeprint[arxiv]{2303.08774}~[cs.CL]


\bibitem[Petridis et~al\mbox{.}(2023)]%
        {anglekindling}
\bibfield{author}{\bibinfo{person}{Savvas Petridis}, \bibinfo{person}{Nicholas Diakopoulos}, \bibinfo{person}{Kevin Crowston}, \bibinfo{person}{Mark Hansen}, \bibinfo{person}{Keren Henderson}, \bibinfo{person}{Stan Jastrzebski}, \bibinfo{person}{Jeffrey~V Nickerson}, {and} \bibinfo{person}{Lydia~B Chilton}.} \bibinfo{year}{2023}\natexlab{}.
\newblock \showarticletitle{Anglekindling: Supporting journalistic angle ideation with large language models}. In \bibinfo{booktitle}{\emph{Proceedings of the 2023 CHI Conference on Human Factors in Computing Systems}}. \bibinfo{pages}{1--16}.
\newblock


\bibitem[Phillips(2009)]%
        {phillips2009film}
\bibfield{author}{\bibinfo{person}{William~H Phillips}.} \bibinfo{year}{2009}\natexlab{}.
\newblock \bibinfo{booktitle}{\emph{Film: an introduction}}.
\newblock \bibinfo{publisher}{Macmillan}.
\newblock


\bibitem[Posner et~al\mbox{.}(2005)]%
        {posner2005circumplex}
\bibfield{author}{\bibinfo{person}{Jonathan Posner}, \bibinfo{person}{James~A Russell}, {and} \bibinfo{person}{Bradley~S Peterson}.} \bibinfo{year}{2005}\natexlab{}.
\newblock \showarticletitle{The circumplex model of affect: An integrative approach to affective neuroscience, cognitive development, and psychopathology}.
\newblock \bibinfo{journal}{\emph{Development and psychopathology}} \bibinfo{volume}{17}, \bibinfo{number}{3} (\bibinfo{year}{2005}), \bibinfo{pages}{715--734}.
\newblock


\bibitem[Rombach et~al\mbox{.}(2021)]%
        {stable_diffusion}
\bibfield{author}{\bibinfo{person}{Robin Rombach}, \bibinfo{person}{Andreas Blattmann}, \bibinfo{person}{Dominik Lorenz}, \bibinfo{person}{Patrick Esser}, {and} \bibinfo{person}{Björn Ommer}.} \bibinfo{year}{2021}\natexlab{}.
\newblock \bibinfo{title}{High-Resolution Image Synthesis with Latent Diffusion Models}.
\newblock
\newblock
\showeprint[arxiv]{2112.10752}~[cs.CV]


\bibitem[Rubin and Agrawala(2014)]%
        {rubin2014generating}
\bibfield{author}{\bibinfo{person}{Steve Rubin} {and} \bibinfo{person}{Maneesh Agrawala}.} \bibinfo{year}{2014}\natexlab{}.
\newblock \showarticletitle{Generating emotionally relevant musical scores for audio stories}. In \bibinfo{booktitle}{\emph{Proceedings of the 27th annual ACM symposium on User interface software and technology}}. \bibinfo{pages}{439--448}.
\newblock


\bibitem[Shi et~al\mbox{.}(2020)]%
        {emog}
\bibfield{author}{\bibinfo{person}{Yang Shi}, \bibinfo{person}{Nan Cao}, \bibinfo{person}{Xiaojuan Ma}, \bibinfo{person}{Siji Chen}, {and} \bibinfo{person}{Pei Liu}.} \bibinfo{year}{2020}\natexlab{}.
\newblock \showarticletitle{EmoG: supporting the sketching of emotional expressions for storyboarding}. In \bibinfo{booktitle}{\emph{Proceedings of the 2020 CHI Conference on Human Factors in Computing Systems}}. \bibinfo{pages}{1--12}.
\newblock


\bibitem[Wang et~al\mbox{.}(2024)]%
        {reelframer}
\bibfield{author}{\bibinfo{person}{Sitong Wang}, \bibinfo{person}{Samia Menon}, \bibinfo{person}{Tao Long}, \bibinfo{person}{Keren Henderson}, \bibinfo{person}{Dingzeyu Li}, \bibinfo{person}{Kevin Crowston}, \bibinfo{person}{Mark Hansen}, \bibinfo{person}{Jeffrey~V. Nickerson}, {and} \bibinfo{person}{Lydia~B. Chilton}.} \bibinfo{year}{2024}\natexlab{}.
\newblock \showarticletitle{ReelFramer: Human-AI Co-Creation for News-to-Video Translation}. In \bibinfo{booktitle}{\emph{Proceedings of the 2024 CHI Conference on Human Factors in Computing Systems}}. \bibinfo{pages}{1--20}.
\newblock


\end{thebibliography}

\appendix
\section{System Prompts}
\subsection{Prompt to Generate Script}
I am making a PSA informing + \textit{audience} + about the problem that + \textit{problem}. 
Additionally, I want to inform them that this problem can be addressed when + \textit{action}. 
Please provide an example description of such a video, making sure to follow key emotional beats that are + \textit{mood}. 
The video should follow a narrative arc, and the narrative goal (for example: Introduction, fostering connection, inspiring action...) of each section should be labeled in under three words. 
Follow this format, and make sure to start every section with *** -  
VISUAL DESCRIPTION: [description of the imagery on screen] TEXT: [text on screen] DURATION: [approximate time of this scene] EMOTIONAL GOAL: [Emotional goal]. 
This video will be made for social media and should be under 45 seconds.

\subsection{Prompt to Regenerate Specific Scene} 
Using this script as context: + \textit{script}, can you replace SCENE + \textit{index} + \textit{scene to be filled} with a scene that achieves the goal of + \textit{scene narrative goal}? 
It should generally have a + \textit{mood} + mood. 
Make sure the text makes sense in the context of the text in the scenes before and after. 
Also, make sure to follow the format of the other parts of the script: 
***TEXT: [onscreen text] IMAGE DESCRIPTION: [image description]. Only return the information for this one scene.

\subsection{Prompt to Generate Art Recommendations} 
\textbf{Image Generation:} 
\textit{art style + color} + illustration of + \textit{image description}

\textbf{Art Style} 
What are words I could use to describe a + \textit{mood} +  mood that is also + \textit{average positivity score to word}? 
For each word, can you provide an art style, movement, or illustration style that is generally representative of that word? 
You don't have to use traditional movements - feel free to be inspired by children's storybook styles, animated styles, styles from advertisements, architecture, and more. 
Start each item of the list with * and follow this format * Word: Style | Explanation. 
For example, entries could be 
* Uplifting: Minimalist Scandinavian Design | typically uses clean lines, neutral color palettes, and natural elements to create a positive and cozy environment that inspires wellness and simplicity 
*Exciting: Action Comic Book illustration in the style of Marvel | this style uses bold and sharp lines, colors, and visual effects to display dynamic action-packed scenes.
Please provide three entries and keep the description under 20 words.

\textbf{Color Palette} 
On a scale of 0-100, 0 meaning ``completely calm'' to 100 meaning ``very excited,'', rank this mood: + \textit{mood}. 
Then, using this assessment, provide a description of colors that could accurately capture this mood. 
For example, for ``completely calm'', you could say ``very muted colors''. 
For ``very excited'', you could say ``very vibrant and saturated colors''. 
Please keep this color description under six words. 
Format your response like this: SCORE: [rank from 0-100] COLOR DESCRIPTION: [color description]" 

\section{Without Mood Prompts (For Technical Evaluation)}
\textbf{Image Generation:} illustration of + \textit{image description}

\textbf{Prompt to Generate Script:}
I am making a PSA informing + \textit{audience} + about the problem that + \textit{problem}. 
Additionally, I want to inform them that this problem can be addressed when \textit{action}.
Please provide an example description of such a video. 
The video should follow a narrative arc, and the narrative goal (for example: Introduction, fostering connection, inspiring action...) of each section should be labeled in under three words. 
Follow this format, and make sure to start every section with *** -  
VISUAL DESCRIPTION: [description of the imagery on screen] TEXT: [text on screen] DURATION: [approximate time of this scene] EMOTIONAL GOAL: [Emotional goal]. This video will be made for social media and should be under 45 seconds.

\section{Messages (For Technical Evaluation)}

\textbf{M1}:
\textit{Audience}: Cat Owners in New York City; 
\textit{Problem}: Free-roaming pet cats are the biggest human-made threat to birds, causing the loss of 2.4 billion birds each year in the US alone;
\textit{Solution}: New Yorkers can help address this issue by keeping their pet cats indoors, and, if allowing them outdoors, keeping them under strict surveillance. \\
\textbf{M2}:
\textit{Audience}: New York Subway Riders; \textit{Problem}: People standing near the doors can create hazards when other passengers enter/leave the train car, resulting in dangerous trips, falls, or other passengers missing their chance to board the train.
\textit{Solution}: New York Subway Riders should move all the way in when they board the train, and move away from the doors to let other passengers on and off. 

\end{document}